\documentclass[conference,12pt,onecolumn]{IEEEtran}

\pdfoutput=1

\usepackage{amsmath,amsfonts}
\usepackage{algorithmic}
\usepackage{algorithm}
\usepackage{array}
\usepackage[caption=false,font=normalsize,labelfont=sf,textfont=sf]{subfig}
\usepackage{textcomp}
\usepackage{stfloats}
\usepackage{url}
\usepackage{verbatim}
\usepackage{graphicx}
\usepackage{cite}
\usepackage[dvipsnames]{xcolor}
\title{Energy Management in a Cooperative Energy Harvesting Wireless Sensor Network}

\author{
    \IEEEauthorblockN{Arghyadeep Barat,  Prabuchandran.K.J, and Shalabh Bhatnagar}

    \thanks{
Arghyadeep Barat is a Bachelor's student in the Department of Electronics and Telecommunication Engineering, Jadavpur University, Kolkata-700032, India. Dr. Prabuchandran K.J. is with the Department of Computer Science and Engineering, Indian Institute of Technology Dharwad-580011, India. Prof. Shalabh Bhatnagar is with the Department of Computer Science and Automation, Indian Institute of Science (IISc), Bangalore-56012, India. (email: arghyadeepbarat123@gmail.com, prabukj@iitdh.ac.in, shalabh@iisc.ac.in).
Arghyadeep Barat was supported through the Summer Research Fellowship from the Indian Academy of Sciences which enabled his visit to IISc for this work. Prabuchandran K.J. was supported by the Science and Engineering Board (SERB), Department of Science and Technology, Government of India for the startup research grant ‘SRG/2021/000048’. Shalabh Bhatnagar was supported by a J.C. Bose Fellowship, Project No. DFTM/02/3125/M/04/AIR-04 from DRDO under DIA-RCOE, a project from DST-ICPS, and the RBCCPS, IISc.
}
}

\begin{document}
\maketitle

\begin{abstract}
In this paper, we consider the problem of finding an optimal energy management policy for a network of sensor nodes capable of harvesting their own energy and sharing it with other nodes in the network. We formulate this problem in the discounted cost Markov decision process framework and obtain good energy-sharing policies using the Deep Deterministic Policy Gradient (DDPG) algorithm. Earlier works have attempted to obtain the optimal energy allocation policy for a single sensor and for multiple sensors arranged on a mote with a single centralized energy buffer. Our algorithms, on the other hand,  provide optimal policies for a distributed network of sensors individually harvesting energy and capable of sharing energy amongst themselves. Through simulations, we illustrate that the policies obtained by our DDPG algorithm using this enhanced network model outperform algorithms that do not share energy or use a centralized energy buffer in the distributed multi-nodal case. 
\end{abstract}

\begin{IEEEkeywords}
Energy Management Policies, Energy Harvesting Wireless Sensor Networks, Cooperative Wireless Sensor Networks, Deep Deterministic Policy Gradient (DDPG) algorithm.
\end{IEEEkeywords}

\section{Introduction}
\IEEEPARstart{E}{nergy} harvesting wireless sensor networks (EHWSNs) are rapidly overshadowing regular WSNs in modern-day IoT applications, for surveillance as well as in monitoring physical and environmental conditions such as temperature, humidity, air pressure, and noise level \cite{Applications}. Sensor nodes require a continuous supply of energy to detect signals and transmit them. The conventional nodes are battery-operated to power the sensors and hence have a finite lifetime which depends on the individual workload. A large enough number of inactive nodes leaves the network inoperable. Since EHWSNs, on the other hand, harvest natural energy, such as solar, thermal, wind, or vibrational energy, given an optimized energy management policy, this energy source can be used as a sustainable limitless powering method for the sensor nodes making their lifetime practically infinite, See \cite{WSN1} and \cite{WSN2}.

Such an energy management policy can be further utilized to distribute energy in microgrids for optimized reallocation of power based on varying rates of energy production and consumption in different centers.

Recent articles such as \cite{EHWSN1, EHWSN2, EHWSN3}, discuss the efficient energy harvesting and utilization mechanisms in order to make EHWSNs a viable alternative. Recent developments in the field of Simultaneous Wireless Information and Power Transfer (SWIPT) have materialized the possibility of developing cooperative sensor networks as well. Technologies are being developed whereby extremely high efficiency \cite{SWIPT, SWIPT1, SWIPT2, SWIPT3}, even up to 90\% is achieved\cite{SWIPT90} in energy transfer using SWIPT at a 2.4GHz frequency range, the highest operating frequency of WSNs as per the IEEE 802.15.4 standard. Therefore, SWIPT has been viably utilized as an available energy-sharing mechanism in a variety of systems such as Distributed Antenna Systems (DAS) \cite{SWIPT_DAS1, SWIPT_DAS2, SWIPT_DAS3}, IoT \cite{SWIPT_IOT1, SWIPT_IOT2, SWIPT_IOT3, SWIPT_IOT4}, WSNs \cite{SWIPT_WSN1, SWIPT_WSN2, SWIPT_WSN3, SWIPT_WSN4} and mobile edge computing \cite{SWIPT_EDGE}. Further, we have articles such as \cite{Coop1, Coop2} further supporting the usage of cooperative WSNs and \cite{CoopWSN1, CoopWSN2, CoopWSN3, CoopWSN4, CoopWSN5} which validate the use of SWIPT technology in wireless communication networks and other such distributed systems. Additionally, \cite{WSNSWIPT1, WSNSWIPT2, WSNSWIPT3, WSNSWIPT4, WSNSWIPT5} refer to the usage of SWIPT for sharing energy to and amongst WSNs and find ways to do so more efficiently.

Earlier work \cite{single} for a single sensor node with finite data and energy buffers utilized the Q-Learning and Speedy Q-Learning Reinforcement Learning (RL) algorithms to optimize the node's performance and optimally manage the energy available. While this provides an optimal Energy Management Policy (EMP) for sensing a transmission in a single node, it is clearly suboptimal in the case of a network of sensors. The primary reason behind this is that nodes in the network with low data influx might overflow with energy while others with high data rates might starve, leading to very high packet loss. Therefore, the harvested energy would not be utilized optimally. 

In \cite{multiple}, a Q-Learning-based algorithm has been proposed for a mote where the energy is harvested and stored in a single central energy buffer and is then distributed among multiple sensors placed on the mote. The model is therefore trained to distribute energy based on the individual data queue length of each node from the central energy buffer. The main reason that \cite{multiple} fails to scale up to decentralized energy harvesting is that it uses a hand-featured value function primarily designed to reduce the growth of state space in a mote.

In our work, we utilize RL algorithms that can automatically learn features using neural networks and further share energy, as shown in \cite{SWIFTWSN2}, between sensors, i.e., from nodes with a high influx of energy to support lower energy nodes leading to more complete utilization of the harvested energy leading to a cooperative approach to reduce the loss of data packets as well as the average latency in transmission.

\subsection{Our Contributions}
\begin{itemize}
 \item We propose a model for Energy Harvesting Wireless Sensor Networks (EHWSNs) that has the ability to share energy amongst sensor nodes for obtaining efficient Energy Management Policies (EMPs).
 \item We formulate an infinite horizon \(\alpha\)-discounted cost minimization problem in the Markov Decision Process (MDP) framework using an appropriate single-stage cost function. 
 \item We solve the MDP for finding an \(\alpha\)-discount optimal EMP using the Deep Deterministic Policy Gradient algorithm because of its ability to handle large state and action spaces thus making our solution scalable
\end{itemize}%

\begin{figure}[htbp]
\centerline{\includegraphics[width=\columnwidth]{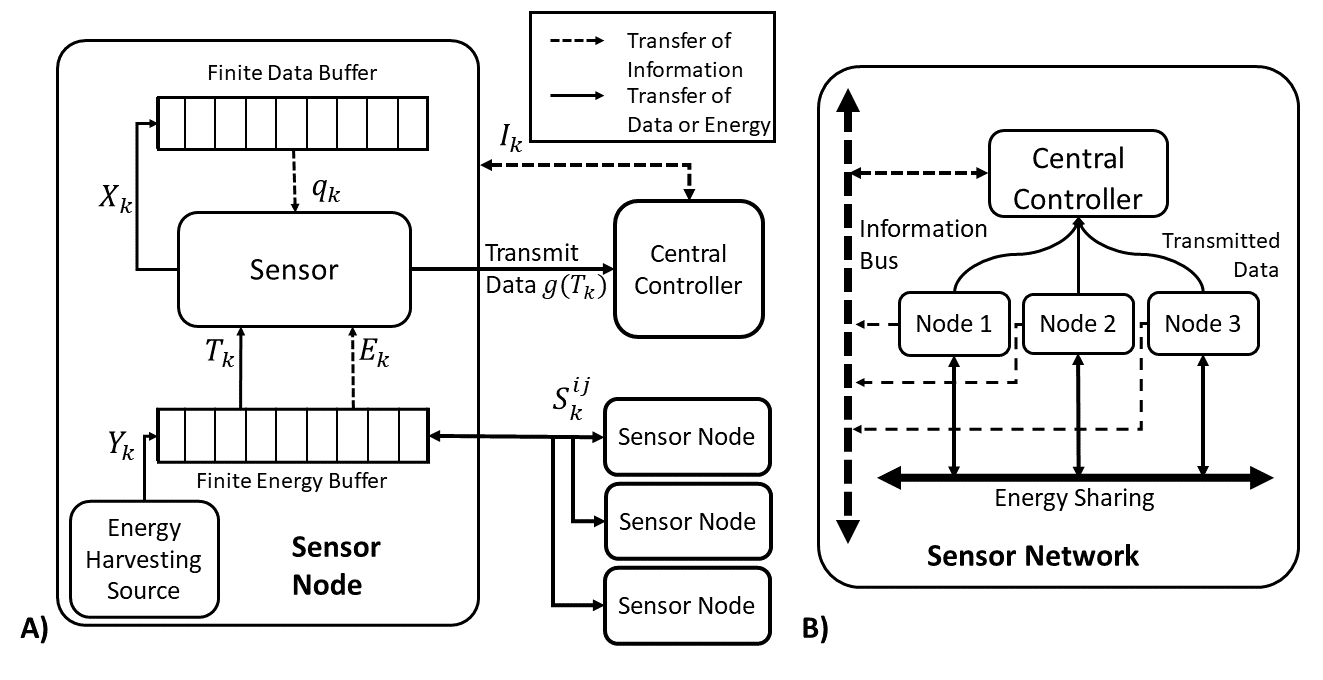}}
\caption{A) Model for an individual sensor node in a sensor network. B) Model for Sensor Network consisting of multiple sensor nodes capable of sharing energy.}
\label{NodeNet}
\end{figure}

\section{Model and Notation}\label{model}
In this section, we describe the model of the energy harvesting sensor network used in our paper. We have considered a discrete time-slotted model for a network consisting of $N$ sensor nodes. We assume the individual sensor nodes have a finite data buffer and a finite energy buffer. The finite buffer assumption is realistic for small-scale sensor nodes. Each of the nodes in the network has its own energy harvesting mechanism from which the energy is stored directly into its own energy buffer. The information regarding the individual data queue and energy levels from all the sensors is sent to a central controller at the end of each slot. The controller determines the energy allocation of each node’s energy for transmission and sharing with every other node and notifies each of the nodes accordingly at the beginning of the next slot. The description of a single sensor node and the network of sensors is depicted in Fig. \ref{NodeNet} (A) and (B) respectively.

In Fig. \ref{NodeNet} (A), the sensor detects a random field and generates corresponding data packets for transmission to the central node. We have assumed a discretized data buffer based on the fact that individual data packets must be transmitted at once at the beginning of a slot avoiding any fractional data packets. The energy buffer however is assumed to be continuous in order to allow maximum efficiency and flexibility for the model to transmit the data as well as share its energy with other sensor nodes. The data and energy buffers have a finite capacity and are denoted by $D_{max}$ (packets or bits) and $E_{max}$ (units of energy) respectively.  In a time slot $k$, the sensor $i$ captures a random field and generates $X_k^i$ units of data.At the same time, $Y_k^i$ units of energy are produced by the energy harvesting mechanism which is stored in the energy buffer. The cumulative state information of the network before the beginning of slot $k$, for sensors $i\in \{1,\dots, N\}$, i.e.,  the queue length of the data buffer $q_{k-1}^i$ and the energy available  $E_{k-1}^i$ denoted by $(q_{k-1}^i, E_{k-1}^i), ~~\forall i\in \{1,\dots, N\}$, is sent to the central controller at the end of the $k-1$st slot. The controller then determines the energy allocation at the next step, $T_k^i$ the amount of energy to be used for transmission of the data packet for node $i$ and $A_k^{ij} ~~\forall j \neq i$ the amount of energy to be shared by the node $i$ to node $j$ at timestep $k$. Note that any amount of energy that node $i$ receives at the timestep $k$ from other nodes will be completely used for transmission in the same slot. 

The conversion function $g(\cdot)$ determines the number of bits that can be transmitted, i.e., if $E$ amount of energy is used, then $g(E)$ bits of data can be transmitted. Therefore, the state variables for $i \in \{1,\ldots,N\}$ can be updated as:

\begin{equation} \label{Queue_Update}
q_{k+1}^i=\left[q_k^i - g \left(T_k^i + \sum_{\substack{j=1 \\ j\neq i}}^{N} A_k^{ji}\right)\right] + X_k^i,
\end{equation}
\begin{equation} \label{Energy_Update}
E_{k+1}^i=E_k^i - T_k^i - \sum_{\substack{j=1\\j\neq i}}^{N} A_k^{ij} + Y_k^i.
\end{equation}
The values of $q_{k+1}^i$ and $E_{k+1}^i$ are then bounded in the ranges $\{0, D_{max}\}$ and $[0, E_{max}]$ respectively. In previous literature \cite{Shannon1, Shannon2, Shannon3, Shannon4}, a logarithmic relation between energy used and data transmitted is assumed based on Shannon's Channel Capacity Theorem. Therefore we have selected a logarithmic conversion function $g(x) = log_2(1+x)$ to define a rather simplistic relation between the energy used and data transmitted while keeping a realistic nonlinear relationship between the two variables as well.

Using equations \eqref{Queue_Update} and \eqref{Energy_Update}, one can simulate the operation of a network of sensor nodes. We assume (A) below on the data and energy arrivals.
\begin{enumerate}
    \item[(A)] $X_k^i$ $(Y_k^i)_{k\geq 1}$, is independent of $\{X_{k-1}^i...X_0^i\}$ $(\{Y_{k-1}^i...Y_0^i\})$ given $q_k^i(E_k^i)$, $T_k^i$ and $A_k^{ij}, i,j \in \{1,\ldots,N\}$. Further, $\{X_k^i\}$ and $\{Y_k^i\}$ are independent of one another for $k\geq 0$. The sequence $\{X_k^i\}_{k\geq 0}$ satisfies $\sup_{k}{E[X_k^i]}\leq r<\infty ~\forall i\in \{1,\ldots,N\}$
\end{enumerate}
Assumption (A) helps establish the Markov property, i.e., all future states of the system are dependent only on the current state and independent of the previous states. We have assumed the random variables corresponding to the amount of data and energy received at each time step to be independent and identically distributed (i.i.d). Both the data arrivals and energy harvesting are modeled as samples from a $Poisson$ distribution with a predetermined mean, as assumed in \cite{single, multiple} and \cite{linear}. The mean is a preset constant assuming that the average rate of energy or data arrival does not change in a small time scale, with all the variability caused because of natural noise. In terms of the energy sharing amongst the nodes, for the simulations in this article, we have assumed the network to be distributed in a small radius so that the efficiency of energy sharing is affected negligibly.

Our central controller consists of the implementation of an RL algorithm that takes information about the states of every node in the network i.e., $(q_k^i, E_k^i)\;\forall i\in \{1,\ldots, N\}$ and recommends $(T_k^i, A_k^{ij})\:\forall i,j \in \{1,\ldots, N\},i\neq j$. Our RL algorithm is based on the actor-critic model wherein the actor proposes the optimal action or energy allocation strategy. The critic model then evaluates the effectiveness of the action by evaluating a corresponding value function. The details of our deep RL model and the algorithm are provided in Section \ref{rl}.

\section{Energy Management Policy via an MDP}\label{mdp}
The Markov Decision Process (MDP) refers to a discrete-time stochastic control process where the actions are chosen in each state so as to minimize some predefined long-term cost. In our problem, the queue length and energy level of each node constitute the state variables. The data queue length $q_k^i \in{0,1,\ldots,D_{max}}$. Energy levels of the nodes are however continuous state variables $E_k^i\in[0, E_{max}],~~\forall i \in \{1,\ldots, N\}, k \geq 0$ where $N$ is the total number of sensor nodes. For the joint state of the network $s_k=(q_k^i, E_k^i),\forall i\in \{1,\ldots,N\}$, the action variables are described as $A_k^{ij}\;\forall i,j\in\{1,\ldots,N\}$. The variable $A_k^{ii}\;\forall i\in(1, N)$ denotes the energy used by the $i^{th}$ node for transmission of data from its own data queue whereas $A_k^{i,j}\;\forall i,j\in\{1,\ldots,N\}, i\neq j$ corresponds to the amount of energy to be shared from the node $i$ to the node $j$ at timestep $k$. The actions determined must follow the energy constraint $ \sum_{j=1}^{N}A_k^{ij} \leq E_{k}^i$. This constraint simply enforces that the nodes can only use a net amount of energy that is bounded by the amount of energy already available for transmission and sharing. A policy $\pi$ is a sequence of maps $A_k$ from the joint state space to the joint action space such that when the joint state is $s_k=(q_k^1, E_k^1,q_k^2, E_k^2,...q_k^N, E_k^N)$ at timestep $k$, $A_k(s_k)$ specifies the energy allocation or distribution for the transmission and sharing amongst each node. By abuse of notation, we denote $A_k(s_k)$ as $A_k$. Therefore, we can denote the joint stationary policy as $\pi = \{A,A,\ldots,A\}$. Based on the assumption (A) stated in Section \ref{model}, we can state that the joint state $\{(q_k^i,E_k^i)\}$ follows the Markov property for all $i \in \{1,\ldots,N\}$ and $\pi\in\Pi$ [the set of all stationary policies].

We define the single-stage cost function as 
\begin{equation}\label{Cost}
    c(q_k^1,\ldots,q_k^N,E_k^1,\ldots,E_k^N,A_k)=\sum_{i=1}^{N}(\Phi(q_k^i)^+),
\end{equation}
where $(q_k^i)^+$ indicates the remaining queue length after the action $A_k^i$ has been chosen and $\Phi$ is any increasing convex function. There are many choices for $\Phi(\cdot)$ like $\Phi(x)=x$, $\Phi(x)=x^2$ or $\Phi(x)= \exp(\alpha x), \alpha \geq 0$. In our experiments, we consider $\Phi(x)=x^2$. A simpler choice for the cost function would be the sum of the queue lengths of each node, in which case, $\Phi(x)=x$. However, such a cost function would minimize the overall sum of queue lengths and will not distinguish between a set of medium queue lengths and a set with highly varying large and small queue lengths. The added benefit of setting the cost function as an increasing convex function such as the sum of squares of queue lengths is that along with minimizing the queue lengths, it tries to reduce the difference in the average queue lengths of the individual sensor nodes.
Therefore we can define the long run $\alpha$-discounted cost $w_\pi(q_0, E_0)$ for a policy $\pi$ as follows:
\begin{equation}\label{Discounted cost}
    w_\pi(q_0, E_0)=E_\pi\left[\sum_{k=0}^{\infty}\alpha^k\cdot(\sum_{i=1}^{N}((q_k^i)^+)^2\mid q_0,E_0)\right]
\end{equation}
where $q_k$ and $E_k$ denote the collective set of queue lengths and energy levels for each node $i$ $(i\in[1,\ldots,N])$ at the $k^{th}$ step and $E_\pi\left[\cdot\right]$ represents the expectation when actions are selected as per policy $\pi$. An $\alpha$-discounted optimal EMP in this setting minimizes $w_\pi(q_0,E_0)$ over all stationary deterministic policies, $\Pi$. Therefore,
\begin{equation}
    w^\star(q_0,E_0) = \min_{\pi\in\Pi}w_\pi(q_0, E_0).
\end{equation}
The primary benefit of representing this problem as an $\alpha$-discounted cost is that a variety of performance objectives can be achieved as per the requirement of the designer through a suitable choice of $\alpha$.

\section{Reinforcement Learning Algorithms}\label{rl}
In this section, we describe the RL algorithms that we have utilized to learn the optimal EMPs for a distributed Energy Harvesting Wireless Sensor Network (EHWSN).

\subsection{\textit{Deep Q Network (DQN)}} The Deep Q Network or DQN model proposed in \cite{DQN} is a neural network model based on the Q-Learning algorithm. The neural network (NN) effectively acts as a function approximation to the Q-table used in Q-Learning \cite{watkins1992q}. Therefore, the weights of the neural network are trained to predict the Q-values of every feasible action associated with a state instead of using the Q-table. Thus, each output value of the DQN model is the Q-value associated with a particular action in that state. 
The optimal action is determined by finding the action having the minimum predicted Q-value. The update rule for training the model involves a gradient descent in the Bellman error loss objective using an NN-based function approximator of the Q-function. The advantage in using the DQN model over Q-Learning is that the number of possible states can be infinite. In our problem, the size of the state space is $(D_{max}\cdot E_{max})^N$ where $D_{max}$ and $E_{max}$ are the maximum capacities of the data buffer and energy buffer, $N$ is the number of nodes, we can see that even if data and energy buffer are both considered discretized, the state space increases exponentially with each additional node. However, since every output node of the DQN corresponds to the Q-value of a unique action, we are still limited by the size of feasible action space in state $s$.

\subsection{\textit{Deep Deterministic Policy Gradient (DDPG)}}
\begin{figure}[htbp]
    \centerline{\includegraphics[width=0.7\columnwidth]{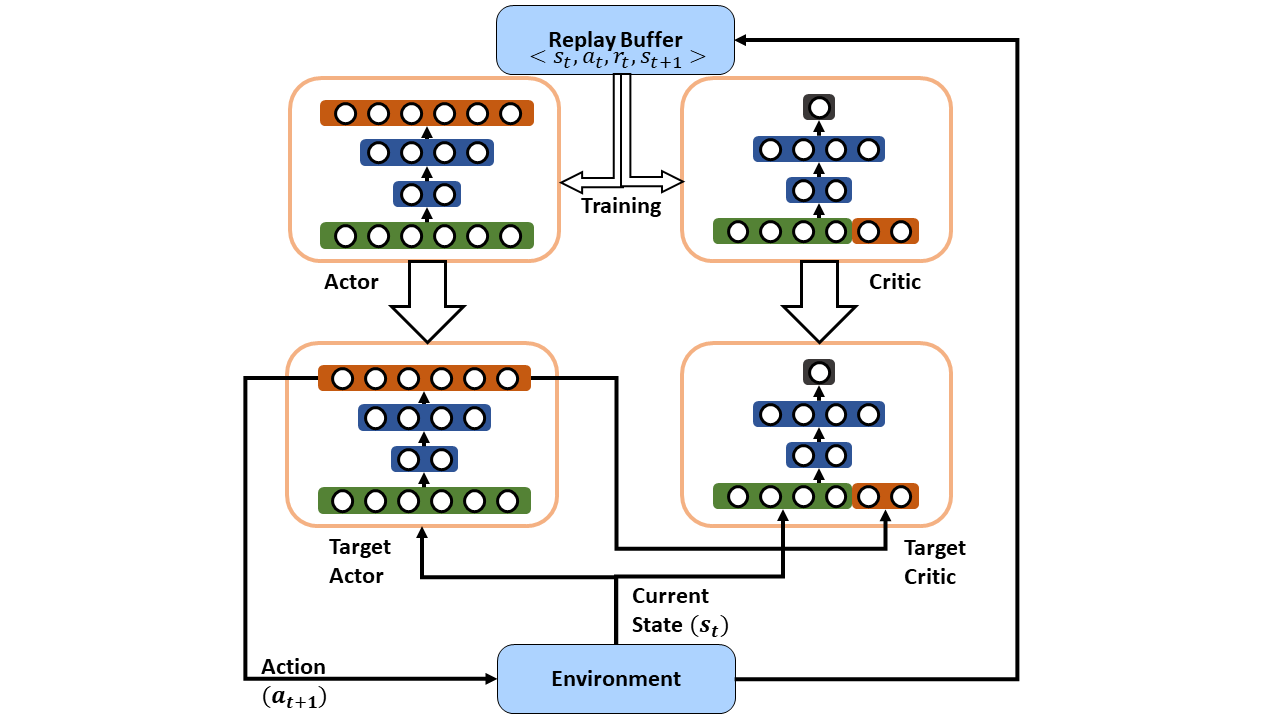}}
    \caption{Training process of the Deep Deterministic Policy Gradient (DDPG) algorithm}
    \label{fig:DDPG}
\end{figure}

The DDPG algorithm proposed in \cite{DDPG} is a model-free actor-critic-based deep RL algorithm. The model used in our case consists of a pair of actor and critic networks as well as a pair of target actor and critic networks as shown in Fig. \ref{fig:DDPG}. The actor models predict the optimal action whereas the critic models evaluate them. The action suggested by the target actor is implemented. Whenever the actor performs better than the target actor, the second network is updated. A representation of the DDPG architecture utilized has been illustrated in Fig. \ref{fig:DDPG}

The biggest advantage of using such a model is the flexibility to operate with both continuous state and action spaces. In our case, although we have taken the data buffer to be quantized, the state space is infinite, since the amount of energy available is a continuous state variable. The actions to be determined are the quantities of energy to be used and shared for each node and hence need to be continuous as well. When the DDPG model starts to learn, the actor network predicts an action to which noise is added in order to explore new actions and evaluate them. The noise function reduces with time as the model converges to the optimal policy.
The critic network predicts the Q-value associated with the action proposed by the actor. When the action is implemented in the environment, the transition is stored in the replay buffer in the form of tuples of the current state, action taken, next state, and reward generated by the environment. The critic loss is generated as a function of the observed reward and the predicted reward. Based on the update of the critic model, the actor model is updated. During training, each time the average performance of the actor and critic model surpasses that of the last recorded best performance, the target actor and critic models are updated with the weights of the trained actor and critic models.

\section{Simulation Results}\label{sim}
The results of our model have been compared to the models given in \cite{single} and \cite{multiple}. In \cite{single}, Q-Learning has been implemented to find the optimal EMP for individual nodes in the network. In \cite{multiple}, a centralized model learns to optimally distribute centrally harvested energy to different nodes in the network utilizing a modified Q-Learning algorithm aided by linear function approximations.  However, in order to cope with larger state and action spaces for comparison, we have implemented it with the DQN algorithm. Finally, our model considers a sensor network capable of sharing energy, the management of which is done using a central controller that learns the optimal policy via the DDPG algorithm.

\textit{Experimental Setup:} In order to compare the different models, we have chosen a setup where the network consists of multiple nodes with varying data rates but identical energy rates with $E[Y_k^i]=5$. The data and energy arrivals $X_k^i, Y_k^i\;\forall k\geq0$ are i.i.d sequences and follow Poisson distribution, i.e., $X_k^i\sim Poisson(E[\lambda_D^i])$ and $Y_k^i\sim Poisson(E[\lambda_E^i])$. We have taken $D_{max}$ and $E_{max}$ to be $10$ each. The conversion function is selected as $g(x)=\log(1+x)$ which determines the amount of data transmitted from the amount of energy used.
\subsection{\textit{Two Nodes Case}}
\begin{figure}[htbp]
    \centering
    \includegraphics[width=0.7\columnwidth]{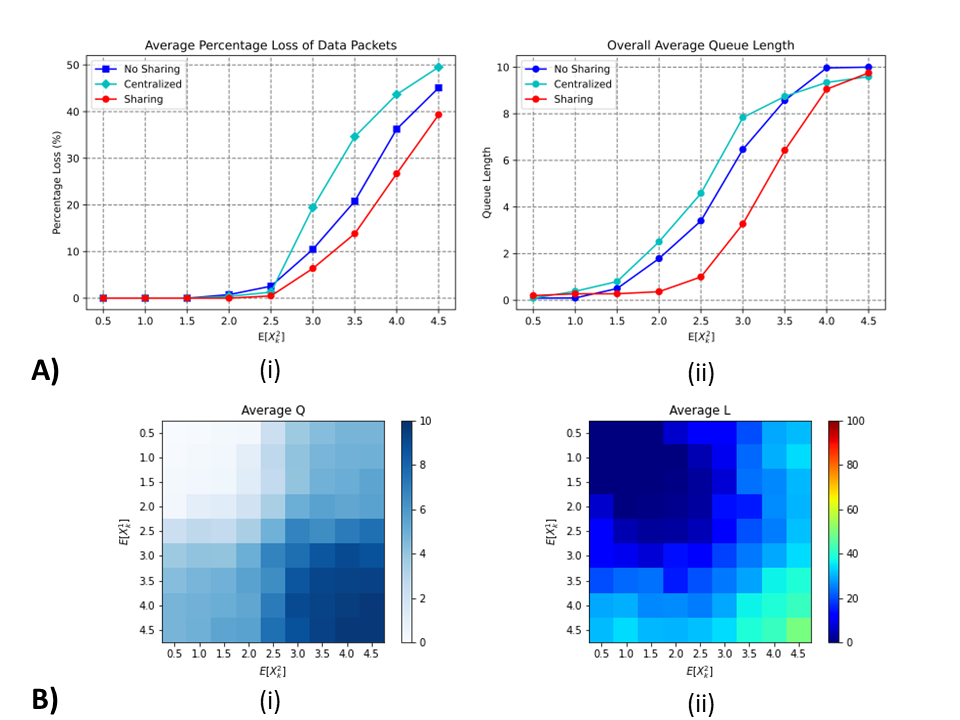}
    \caption{Comparison with earlier models \cite{single}, \cite{multiple} in terms of Average Queue Length [A(i)] and Average Percentage Loss of Data Packets of Network [A(ii)] and Heat-map of Queue Length [B(i)] and Data Loss Percentage [B(ii)] for optimal policy learned by the current model. $X_k^i, Y_k^i \sim Poisson Distribution$; $E[Y_k^i]=5\;\forall i ~\in {1,2}$ with the condition of $E[X_k^1]=0.5$ assumed for A(i) and A(ii)}
    \label{fig:Combo}
\end{figure}

 In order to highlight the advantages of sharing energy, as considered in our model, the average data rate for one node is set at $E[X_k^1]=0.5$. The performance of the network is measured using two parameters, the long-run average queue length and the average percentage loss of data packets measured across different average data arrival rates for the second node. Therefore, $E[X_k^2]$ is plotted along the X-axis and for the two plots in Fig. \ref{fig:Combo}A(i) and (ii) long-run average queue length and data packet loss are plotted along the Y-axis respectively as an additional performance metric. In the following figure, the models proposed in \cite{single} and \cite{multiple} have been referred to as the ``No Sharing" and the ``Centralized" model respectively whereas our model has been referred to as the ``Sharing model".

Fig. \ref{fig:Combo}A(i) illustrates the fact that compared to the models proposed in \cite{single} and \cite{multiple}, our decentralized model converges to policies having lower average data queue length. Based on Little's Law \cite{Little1}, we can state that the long-term average queue length is equal to the product of the long-term average data arrival rate and the average time delay for transmission. Therefore, we can directly conclude, that the policies derived by using our model lead to a lower average time delay in the transmission of the data (being collected) by the network as a whole.

Fig. \ref{fig:Combo}A(ii) shows that our model successfully develops an energy split profile for sharing amongst neighboring nodes in order to effectively distribute the varying load of data arrival rates in different nodes. The model presented in \cite{multiple}, although more cost-effective, compromises the accuracy of convergence to the optimal policy primarily due to linear function approximation and limitations to the energy distribution mechanism. This leads to a slightly higher average queue length, hence a higher transmission delay as well as a higher loss of data packets. In clear contrast, our model successfully finds a policy to share energy amongst neighboring nodes in a sensor network to efficiently reduce the transmission delay as well as the loss of data packets beyond the possible margins for nodes operating individually. A correlated equilibrium is established by individual sensor nodes cooperating for the collective gain of the sensor network. Fig. \ref{fig:Combo}B(i) and (ii) demonstrate the variation of long-run average queue length and percentage loss of data packets for a network consisting of two nodes for mean data arrival rate varying from 0.5 to 4.5 for each node, with an interval of 0.5.

Since the expected energy arrival rate has been fixed for each node, a critical data arrival rate can be calculated as

\begin{equation}
    E[T_k] = E[g(\sum_{i=1}^{N} E[Y_k^i])].
\end{equation}

Here, $T_k$ is the maximum average rate of data that can be transmitted given the current distribution of $Y_k^i$. Hence it is the same as the critical rate or the maximum average rate at which data arrival can be handled by the arriving energy. Therefore, beyond this capacity, the network should become unstable leading to very high queue length as well as a high percentage of packet loss. For a network consisting of two nodes, each receiving energy at a mean rate of 5, the critical rate for the network as a whole comes out to be 3.395 approximately. Similarly in the results corresponding to our model as presented Fig. \ref{fig:Combo}, we observe a large increase in both beyond the point where $\sum_{i=1}^{N} E[X_k^i] \geq E[T_k]$. Even then, due to the ability of the model to share energy amongst nodes, it reduces the queue length and packet loss much lower than what is achieved using previously proposed models \cite{single}, \cite{multiple}. The optimal policies derived by our model are such that even when the data arrival rate is about 2.7 times the critical rate, i.e., the maximum rate at which the network is stable, the percentage of data loss is limited to 43\%.

\subsection{\textit{Scalability}}
One of the added benefits of using Deep RL algorithms such as DDPG is the possibility to solve for larger state and action spaces, hence allowing for the scalability of our model. We have simulated each of the aforementioned algorithms, i.e., the Q-Learning-based No Sharing model, the DQN-based Centralized model, and our DDPG-based Energy Sharing model. In each case, simulations have been carried out for networks containing multiple nodes, each with $E[Y_k^i]=5$ while $E[X_k^i]$ is randomly selected in the range of 0 to 4.

For a simulation having 10 nodes, the minimum percentage of data loss is given as 43\%, 17\%, and 11\% for each of the models respectively. Therefore, our model is able to handle the same load of data influx with the lowest data loss rate among the three algorithms.

\begin{figure}[htbp]
    \centering
    \includegraphics[width=0.7\columnwidth]{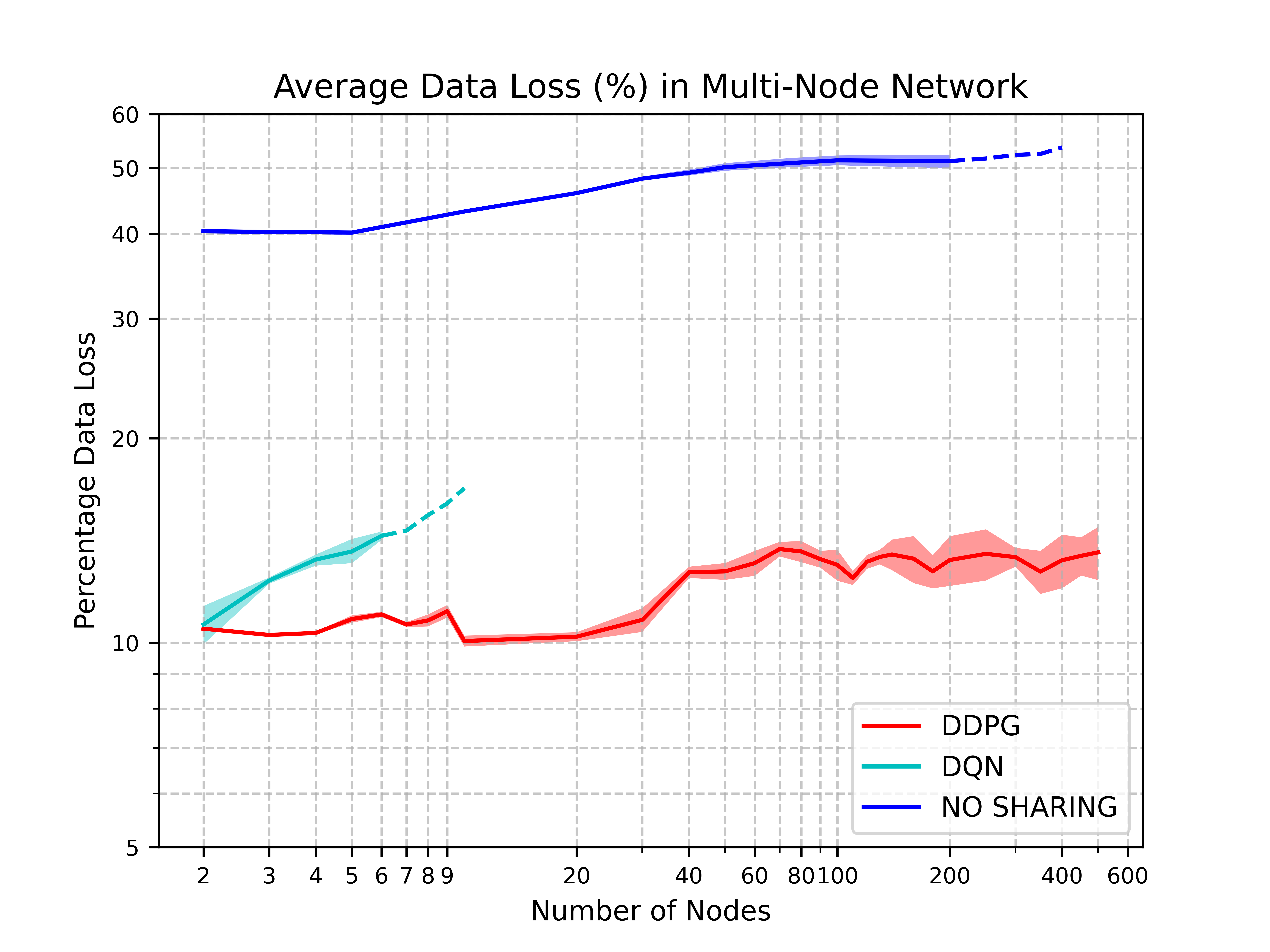}
    \caption{Comparison with earlier models \cite{single}, \cite{multiple} in terms of Average Percentage Data Loss for optimal policy learned by the respective models. $E[Y_k^i]=5\;\forall i ~\in \{1,\ldots,N\}$ ($N$=number of sensor nodes in network) \; $E[X_k^i]$ is randomly selected $\forall i ~\in \{1,\ldots,N\}$ such that the range is $[0,4]$ and the mean is $2$.}
    \label{fig:Scale}
\end{figure}

Now, for simulations in the same device, the largest size of a network possible turns out to be 200 nodes, 6 nodes, and 500 nodes for each algorithm respectively. The first model is restricted by the amount of  RAM to process the Q-Table and the space required for storing the entire table. In the second model using the DQN model, every possible combined action has to be considered as an output. For our model, neural networks with just two hidden layers having only 2 and 4 units respectively for the actor and critic networks have been used. Using the same neural architecture, we can reliably solve the optimization problem for even up to 500 nodes. Larger simulations are restricted by RAM. Fig. \ref{fig:Scale} demonstrates the superior performance of our model compared to the other two models. All the above simulations are done using the software Virtual Studio Code $v 1.75.1$ on an Intel $i5-10210U$ processor computer with a clock speed of 1.60 GHz. 

Fig. \ref{fig:Scale} shows the average percentage of data loss due to the policies learned by the compared algorithms and their variability over multiple runs. The solid line represents the average data loss results obtained over multiple runs on the same device. The dotted lines show the extension of the results with additional computational resources. The results clearly demonstrate that with the same amount of computational resources, our model is able to optimize to a better result even for a much larger network. 

In our model, in contrast to the others in comparison, the policies learned are almost equally optimal with data loss increasing to only 13\% in the 500-node network. The average data and energy queue levels maintained at almost every node in the 500-node network were approximately identical to the same for the optimized policy in a network of 2 nodes. Therefore, it can outperform the other methods even for much larger networks. The above results demonstrate the scalability and further optimality of our model in comparison with the other models described above.

\section{Implementation Details}
The model architecture consists of the DDPG model which has a pair of training actor and critic networks and a pair of target actor and critic networks. Each actor and critic network has an identical internal structure of just two hidden layers with two units on the first layer and 4 units on the second. This makes for a very light model allowing for scalability. The output layer for the actor-network consists of $N^2$ units, $N$ is the number of nodes in the network. This is because every node corresponds to $N$ action variables since it decides the amount of energy to be used for transmission in order to clear its own data queue and the amount of energy to be shared individually with the other $N-1$ nodes in the network. The energy left after subtracting the sum from the energy level is the amount stored for future use of the node itself. We have been able to derive our results with such a light architecture because of its ability to generalize a low-level policy learned for energy distribution in smaller WSNs to a much larger scale with equal efficiency.

\section{Conclusion and Future Work}\label{conc}
We formulated the problem of energy sharing and distribution in Energy Harvesting Wireless Sensor Networks (EHWSNs) as Markov Decision Processes (MDP) and studied an application of a Deep Deterministic Policy Gradient algorithm to find the optimal policy for minimizing the transmission delay for the sensor network. Owing to the energy-sharing capability of the network and due to the efficient energy usage policy, our model succeeds in minimizing the loss of data packets in case of an overload of individual nodes by rerouting the energy harvested from other nodes. The usage of the DDPG algorithm enables learning the optimal policy without an explicit model of the environment and also improves the joint state and action space handling capacity of the algorithm. The benefits of the proposed model and used algorithm have been established via experimental results and simulation that demonstrate significantly lower average queue length, transmission delay, and percentage loss of data in overloaded situations. The results can also be reproduced with limited computational resources at a much larger scale compared to the compared algorithms

The current model becomes computationally inefficient on a larger scale due to the rapidly increasing computational cost as well as the time required for training the model for learning policies over larger state and action spaces. Hence, models can be developed to create a layered structure to classify the network into clusters to reduce individual computational costs as well as parallelize the learning of optimal policies for different clusters.

In the future, we would like to extend the energy distribution protocol so that nodes are classified into smaller clusters and only members of the same cluster can share energy with each other. This would also help with the decentralization of the control. Furthermore, we would like to incorporate intricacies related to the efficiency of energy sharing like loss in efficiency due to wireless transfer and the variability of average data arrival or energy arrival rate in our future work.

\bibliographystyle{ieeetr}

\bibliography{References} 

\newpage

\vfill

\end{document}